\documentclass{llncs}
\pdfoutput=1
\usepackage[pdftex]{graphicx}
\usepackage{color}
\usepackage{listings}
\usepackage{array}

\graphicspath{{figures/}}
\DeclareGraphicsExtensions{.pdf,.png}

\definecolor{somegrey}{rgb}{.95,.95,.99}
\lstdefinelanguage{ASM}{
morekeywords = { addpd, mulpd, addsd, subsd,
movsd, movdqa, movlhps, movhlps,
dec, jnz, mov, movdq, prefetchnta,
add, sub, movlpd, movhpd, xor, movapd,
movaps, add, cmp, jb, js,
pop, push, movntpd, div},
morecomment=[l];,
}

\lstnewenvironment{ASMSource}[1][]{
\lstset{#1}
\lstset{frame=tb}
\lstset{numbers=left, numberstyle=\tiny, stepnumber=1}
\lstset{backgroundcolor=\color{somegrey}}
\lstset{language=ASM}
}{
}

\lstnewenvironment{CSource}[1][]{
\lstset{#1}
\lstset{frame=tb}
\lstset{numbers=left, numberstyle=\tiny, stepnumber=1}
\lstset{backgroundcolor=\color{somegrey}}
\lstset{language=C}
}{
}
\begin{document}
\mainmatter              
\title{Introducing a Performance Model for Bandwidth-Limited Loop Kernels}
\titlerunning{Performance Model}  
%
\author{Jan Treibig \and Georg Hager}
\authorrunning{Jan Treibig et al.}
\institute{Regionales Rechenzentrum Erlangen\\ University Erlangen-Nuernberg\\
\email{jan.treibig@rrze.uni-erlangen.de}}

\maketitle
\begin{abstract}
    We present a performance model for bandwidth limited loop kernels which is
    founded on the analysis of modern cache based microarchitectures. This model
    allows an accurate performance prediction and evaluation for
    existing instruction codes. It provides an in-depth understanding of how
    performance for different memory hierarchy levels is made up. The
    performance of raw memory load, store and copy operations and a stream
    vector triad are analyzed and benchmarked on three modern x86-type quad-core
    architectures in order to demonstrate the capabilities of the model.
\end{abstract}
%
\section{Introduction}
%
Many algorithms are limited by bandwidth, meaning that the memory subsystem
cannot provide the data as fast as the arithmetic core could process it. One
solution to this problem is to introduce multi-level memory hierarchies with
low-latency and high-bandwidth caches which exploit temporal locality in an
application's data access pattern. In many scientific algorithms the bandwidth
bottleneck is still severe, however. While there exist many models predicting
the influence of main memory bandwidth on the performance \cite{schoenauer}, less is known about
bandwidth-limited in-cache performance. Caches are often assumed to be
infinitely fast in comparison to main memory. Our proposed model explains what
parts contribute to the runtime of bandwidth-limited algorithms on all memory
levels. We will show that meaningful predictions can only be drawn if the
execution of the instruction code is taken into account.

To introduce and evaluate the model, basic building blocks of streaming
algorithms (load, store and copy operations) are analyzed and benchmarked on
three x86-type test machines. In addition, as a prototype for many streaming
algorithms we use the STREAM triad $\vec{A}=\vec{B}+\alpha * \vec{C}$, which
matches the performance characteristics of many real algorithms \cite{jalby}. The main
routine and utility modules are implemented in C while the actual loop code
uses assembly language. The runtime is measured in clock cycles using the
\verb+rdtsc+ instruction.

Section \ref{sec:machines} presents the microarchitectures and technical
specifications of the test machines. In Section~\ref{sec:model} the model
approach is briefly described. The application of the model and according
measurements can be found in Sections \ref{sec:theory} and \ref{sec:results}.
%
\section{Experimental test-bed}
\label{sec:machines}
\begin{figure}[b]
  \centering
    \includegraphics[width=0.31\linewidth]{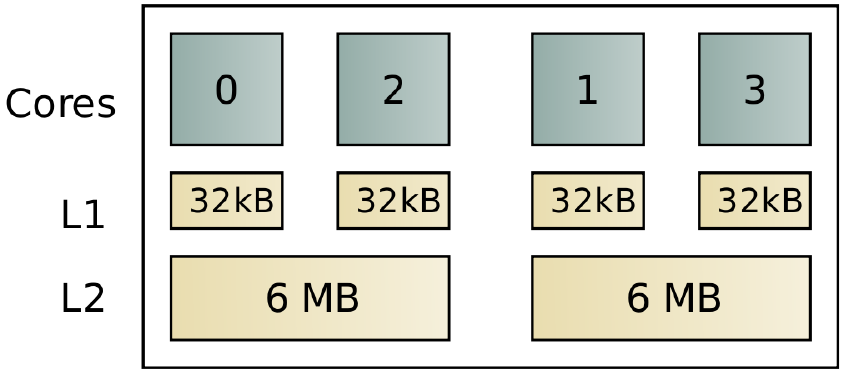}
    \includegraphics[width=0.31\linewidth]{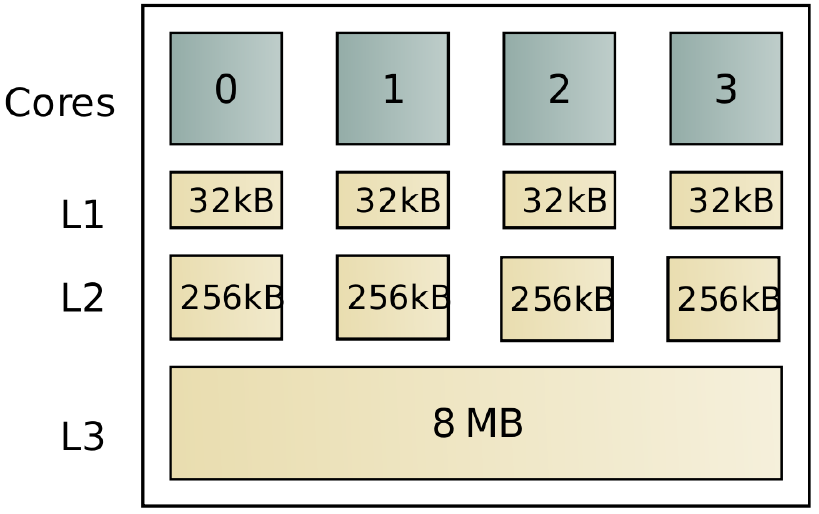}
    \includegraphics[width=0.31\linewidth]{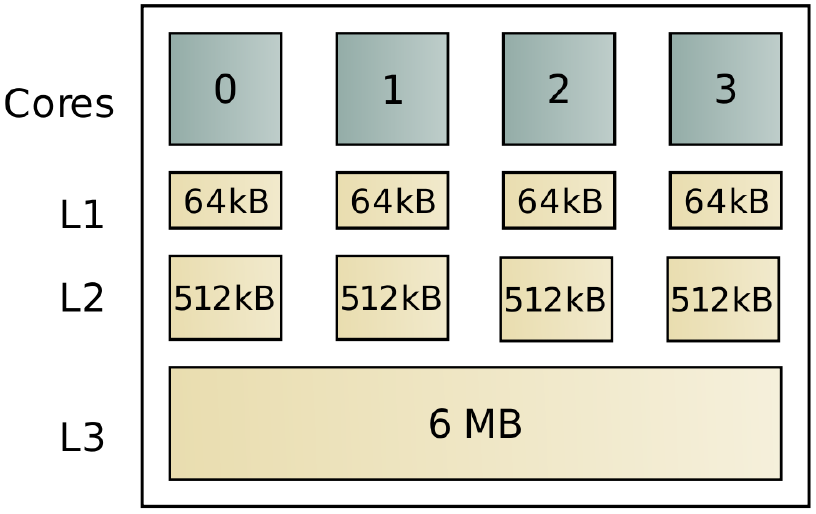}
    \caption{Cache group structure of the multi-core architectures in the
    test-bed for Core 2 (left), Core i7 (middle) and Shanghai (right)}
    \label{fig:cache_arch}
\end{figure}

%
An overview of the test machines can be found in Table
\ref{tab:arch}. As representatives of current x86
architectures we have chosen Intel ``Core 2 Quad'' and ``Core i7''
processors, and an AMD ``Shanghai'' chip. The cache group structure, 
i.e., which cores share caches of what size, is
illustrated in Figure \ref{fig:cache_arch}. For detailed information
about microarchitecture and cache organization, see the
Intel \cite{IntelOpt} and AMD \cite{AMD_SW_Opt:07} Optimization Handbooks.
Although the Shanghai processor used for the tests sits in a dual-socket
motherboard, we restrict our analysis to a single core.
\begin{table}[tb]
    \caption{Test machine specifications. The cache line size is 64 bytes 
	for all processors and cache levels.}
    \label{tab:arch}
    \centering
	\begin{tabular}{lccc}
	    \hline
	    &Core 2&Nehalem&Shanghai\\
	    &Intel Core2 Q9550&Intel i7 920&AMD Opteron 2378 \\
	    \hline
	    Execution Core&&\\
	    \hline
	    Clock [GHz]&2.83&2.67&2.4\\
	    Throughput&4 ops&4 ops&3 ops\\
	    Peak FP rate MultAdd&4 flops/cycle&4 flops/cycle&4 flops/cycle\\
	    \hline
	    L1 Cache&32 kB     &32 kB     &64 kB \\
	    Parallelism     &4 banks, dual ported &4 banks, dual ported&8 banks, dual ported\\
	    \hline
	    L2 Cache&2x6 MB (inclusive)&4x256 KB &4x512 KB (exclusive)\\
	    \hline
	    L3 Cache (shared)&-&8 MB (inclusive)&6 MB (exclusive)\\
	    \hline
	    Main Memory&DDR2-800&DDR3-1066&DDR2-800\\
	    Channels&2&3&2\\
	    Memory clock [MHz]&800&1066&800\\
	    Bytes/ clock&16&24&16\\
	    Bandwidth [GB/s]&12,8&25,6&12,8\\
	    \hline
	\end{tabular}
\end{table}
%
\section{Performance Model}
\label{sec:model}
%
This model proposes an iterative approach to analytically predict the
performance of bandwidth-limited algorithms in all memory hierarchy levels. The
basic building block of a streaming algorithm is its computational kernel in
the inner loop body. The kernel is performance-limited by the L1 cache, i.e.\
the maximum number of load and store accesses per cycle, and the capability of
the pipelined, superscalar core to execute instructions. All lower levels of
the memory hierarchy are reduced to their bandwidth properties, with data paths
and transfer volumes based on the real cache architecture.
The minimum transfer size between memory levels is one cache line.
Based on the transfer volumes and the bandwidth capabilities, the
contributed cycles of each transfer are summed up with the cycles needed to
execute the instructions with data coming from L1 cache. The result is the time
needed to execute the loop kernel, assuming there is no access latency or
overlap of contributions. 

It must be stressed that a correct application of this model requires intimate
knowledge of cache architectures and data paths. This information is available
from processor manufacturers \cite{IntelOpt,AMD_SW_Opt:07}, but sometimes the
level of detail is insufficient for fixing all parameters and relevant
information must be derived from measurements.

%
%
%
\section{Theoretical Analysis}
\label{sec:theory}
%
In this section we predict performance numbers for each benchmark on
all memory levels, based on an architectural analysis of the
processors used in the test-bed. Unless otherwise noted, all results
are given in CPU cycles.

As mentioned earlier, basic data operations in L1 cache are limited by cache
bandwidth, which is determined by the load and store instructions that can
execute per cycle. The Intel cores can retire one 128-bit load and one 128-bit
store in every cycle. L1 bandwidth is thus limited to 16 bytes per cycle if
only loads (or stores) are used, and reaches its peak of 32 bytes per cycle
only for a copy operation. The AMD Shanghai core can perform either two
128-bit loads or two 64-bit stores in every cycle. This results in a load
performance of 32 bytes per cycle and a store performance of 16 bytes per
cycle. 

For load-only and store-only kernels, there is only one data stream,
i.e., exactly one cache line is processed at any time. With copy
and stream triad kernels, this number increases to two and three,
respectively. Together with the execution limits described above it
is possible to predict the number of cycles needed to execute the instructions
necessary to process one cache line per stream
(see the ``L1'' columns in Table~\ref{tab:cache_model})\@.
\begin{table}[tb]
    \caption{Theoretical prediction of execution times for eight loop
	iterations (one cache line per stream) on Core 2 (A), Core i7 (B), and
	Shanghai (C) processors}
    \label{tab:cache_model}
    \centering\renewcommand{\arraystretch}{1.15}\addtolength{\tabcolsep}{1mm}
\begin{tabular}{r|ccc|ccc|cc|ccc}
          &\multicolumn{3}{c|}{L1}&\multicolumn{3}{c|}{L2}&\multicolumn{2}{c|}{L3}&\multicolumn{3}{c}{Memory}\\
          &A  &B  &C              &A   &B   &C            &B   &C                 &A     &B      &C    \\
    \hline
    Load  &4  &4  &2              &6   &6   &6            &8   &8                 &20    &15     &18   \\
    Store &4  &4  &4              &8   &8   &8            &12  &10                &36    &26     &32   \\
    Copy  &4  &4  &6              &10  &10  &14           &16  &18                &52    &36     &50   \\
    Triad &8  &8  &8              &16  &16  &20           &24  &26                &72    &51     &68   \\
\end{tabular}
\end{table}

L2 cache bandwidth is influenced by three factors: (i) the finite bus width
between L1 and L2 cache for refills and evictions, (ii) the fact that
\emph{either} ALU access \emph{or} cache refill can occur at any one time, and
(iii) the L2 cache access latency. All three architectures have a 256-bit bus
connection between L1 and L2 cache and use a write back and write allocate
strategy for stores. In case of an L1 store miss, the cache line is first moved
from L2 to L1 before it can be updated (write allocate)\@. Together with its
later eviction to L2, this results in an effective bandwidth requirement of 128
byte per cache line write miss update. 

On the Intel processors, a load miss incurs only a single cache line transfer
from L2 to L1 because the cache hierarchy is inclusive. The Core i7 L2 cache is
not strictly inclusive, but for the benchmarks covered here (no cache line sharing
and no reuse) an inclusive behavior was assumed due to the lack of detailed
documentation about the L2 cache. In contrast, the AMD L2 cache is exclusive:
It only contains data that was evicted from L1 due to conflict misses. On a
load miss the new cache line and the replaced cache line have to be exchanged.
This results in a bandwidth requirement of two cache lines for every cache line
load from L2. 

The overall execution time of the loop kernel on one cache line per
stream is the sum of (i) the time needed to transfer the cache line(s)
between L2 and L1 and (ii) the runtime of the loop kernel in L1 cache.
Table~\ref{tab:transfersum} shows the different contributions for pure
load, pure store, copy and triad operations on Intel and AMD
processors. Looking at, e.g., the copy operation on Intel, the model
predicts that only 6 cycles out of 10 can be used to transfer data
from L2 to L1 cache. The remaining 4 cycles are spent with the
execution of the loop kernel in L1. This explains the well-known
performance breakdown for streaming kernels when data does not fit
into L1 any more, although the nominal L1 and L2 bandwidths are
identical.
\begin{table}[tb]
    \caption{Loop kernel runtime for one cache line per stream in L2 cache}
    \label{tab:transfersum}
    \centering\addtolength{\tabcolsep}{1mm}
\begin{tabular}{r|cccc|cccc}
    &\multicolumn{4}{c|}{Intel}&\multicolumn{4}{c}{AMD}\\
       &Load&Store&Copy&Triad&Load&Store&Copy&Triad\\
    \hline
    L1 part &4   &4    &4   &8   &2   &4    &6   &8\\
    L2 part &2   &4    &6   &8   &4   &4    &8   &12\\
    \hline
    L1+L2   &6   &8    &10  &16  &6   &8    &14  &20\\
\end{tabular}
\end{table}
All results are included in the ``L2'' columns of
Table~\ref{tab:cache_model}\@. The large number of cycles for the AMD
architecture can be attributed to the exclusive cache structure, which leads to
a lot of additional inter-cache traffic.

Not much is known about the L3 cache architecture on Intel Core i7 and AMD
Shanghai. It can be assumed that the bus width between the caches is 256 bits,
which was confirmed by our measurements. Our model assumes a strictly
inclusive cache hierarchy for the Intel designs, in which L3 cache is ``just
another level.'' For the AMD chips it is known that all caches share a single
bus. On an L1 miss, data is directly loaded into L1 cache. Only evicted L1
lines will be stored to lower hierarchy levels. While the L3 cache is not
strictly exclusive on the AMD Shanghai, exclusive behavior can be assumed
for the benchmarks considered here.
Under these assumptions, the model can predict the required number of cycles in
the same way as for the L2 case above. The ``L3'' columns in
Table~\ref{tab:cache_model} show the results.

\begin{figure}[tb]\centering
    \includegraphics[width=0.9\linewidth]{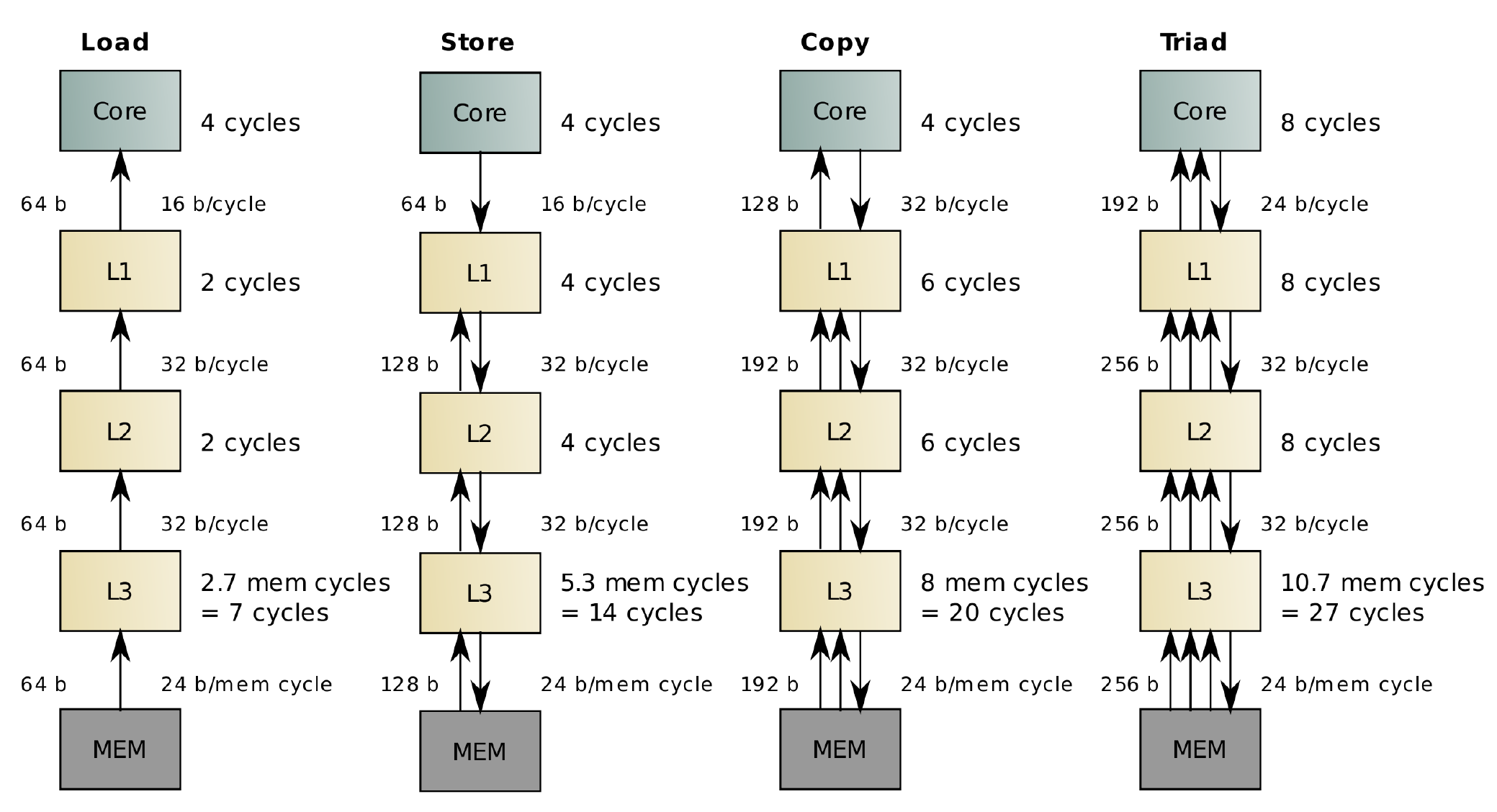}
    \caption{Main memory performance model for Intel Core i7.
    There are separate buses connecting the different cache levels.}
    \label{fig:mem_data_intel_nehalem}
\end{figure}

\begin{figure}[tb]\centering
    \includegraphics[width=0.9\linewidth]{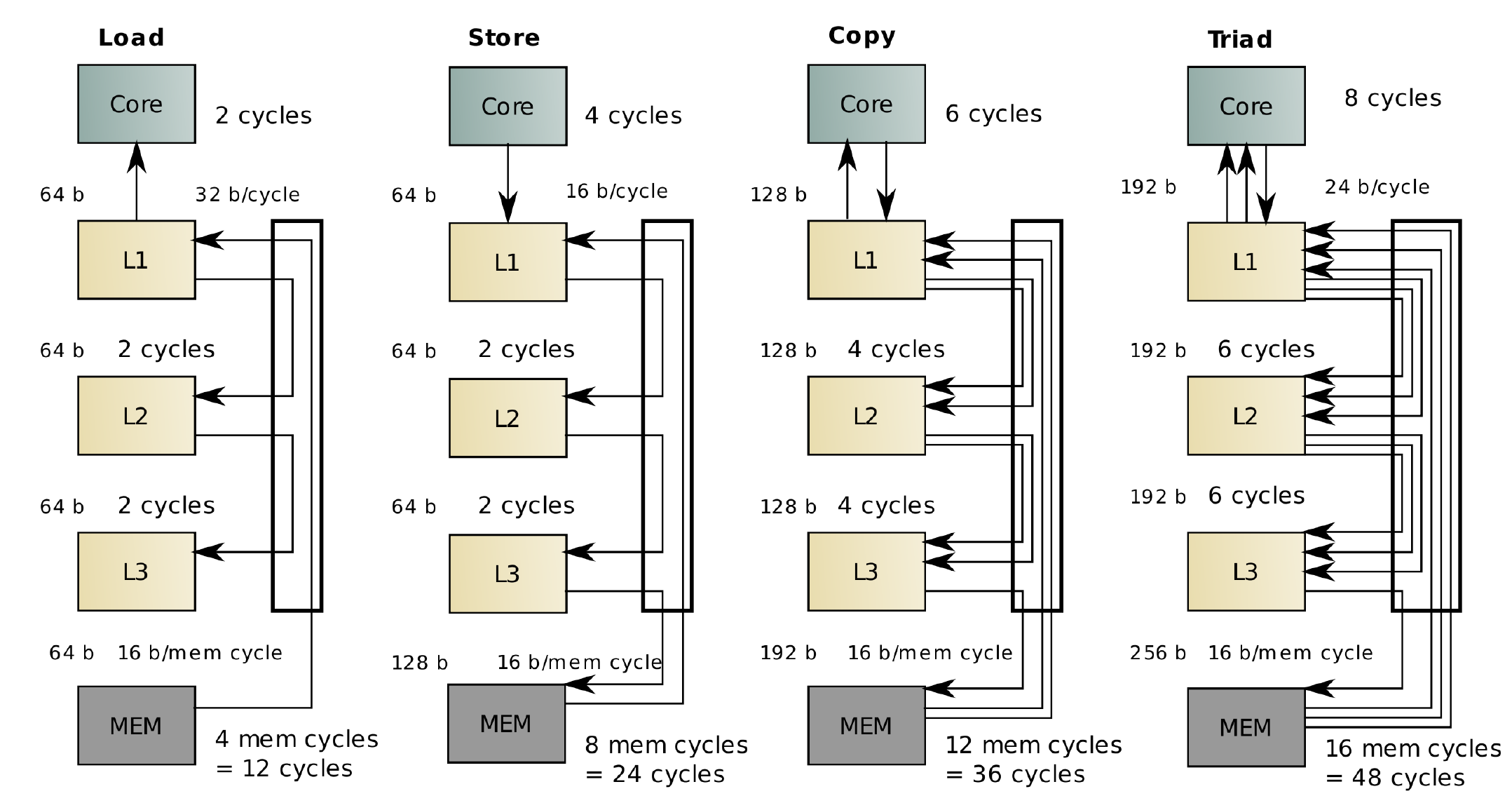}
    \caption{Main memory performance model for AMD Shanghai. All 
    caches are connected via a shared bus.}
    \label{fig:mem_data_amd_Shanghai}
\end{figure}

If data resides in main memory, we again assume a strictly hierarchical
(inclusive) data load on Intel processors, while data is loaded directly into
L1 cache on AMD even on store misses.
The cycles for main memory transfers are computed using the effective memory
clock and bus width and are converted into CPU cycles. For consistency reasons, 
non-temporal (``steaming'') stores were not used for the main memory regime.
Data transfer volumes and rates, and predicted cycles for a cache line update
are illustrated in Figures~\ref{fig:mem_data_intel_nehalem} (Core i7) and
\ref{fig:mem_data_amd_Shanghai} (Shanghai). They are also included in the
``Memory'' columns of Table~\ref{tab:cache_model}\@.
%
\section{Measurements}
\label{sec:results}
%
Measured cycles for a cache line update, the ratio of predicted versus measured
cycles, and the real and effective bandwidths are listed in
Table~\ref{tab:cache_results}. Here, ``effective bandwidth'' means the
bandwidth available to the application, whereas ``real bandwidth'' refers to
the actual data transfer taking place. For every layer in the hierarchy the
working set size was chosen to fit into the appropriate level, but not into
higher ones. The measurements confirm the predictions of the model well in the
L1 regime, with slightly larger deviations for the AMD architecture. This
might be caused either by non-optimal code or the maximum throughput of three
macro-ops per cycle, which is at its limit in these benchmarks (the Intel
designs allow up to four macro-ops per cycle)\@. In general, we refer to
additional (measured) cycles spent compared to the model as ``overhead.''

Also the L2 results confirm the predictions. One exception is the store
performance of the Intel Core i7, which is significantly better than the
prediction. This indicates that the model does not describe the store behavior
correctly. At the moment we have no additional information about the L2
behavior on Core i7 to solve this problem. The overhead for accessing the L2
cache with a streaming data access pattern scales with the number of involved
cache lines, as can be derived from a comparison of the measured cache line
update cycles in Table~\ref{tab:cache_results} and the predictions in 
Table~\ref{tab:cache_model}\@. The highest cost occurs on the Core 2 with 2 cycles per cache line
for the triad, followed by Shanghai with 1.5 cycles per cache line. Core i7 has
a very low L2 access overhead of 0.5 cycles per cache line. 
Still, all Core i7 results must be interpreted with caution
until the L2 behavior can be predicted correctly by a revised model. 
All architectures are good at hiding cache latencies for
streaming patterns.

On the AMD Shanghai there is significant overhead involved in accessing the L3
cache. On Core i7 the behavior is similar to the L2 results: The store result
is better than the prediction, which influences all other test cases involving
a store. It is obvious that the Core i7 applies an unknown optimization for
write allocate operations. The effective bandwidth available to the
application is dramatically higher on the Intel Core i7 owing to the inclusive
cache hierarchy, while the AMD Shanghai works efficiently within the limits of
its architecture but suffers from a lot of additional cache traffic due to its
exclusive caches.
\begin{table}[tb]
    \caption{Benchmark results}
    \label{tab:cache_results}
    \centering\footnotesize
	\begin{tabular}{r|cccc|cccc|cccc|cccc}
	               &\multicolumn{4}{c|}{L1}       &\multicolumn{4}{c|}{L2}        &\multicolumn{4}{c}{L3}           &\multicolumn{4}{c}{Memory}        \\
	               &Load  &Store  &Copy   &Triad &Load   &Store  &Copy   &Triad &Load   &Store   &Copy   &Triad  &Load     &Store    &Copy       &Triad  \\
	    \hline                                                                                                                                                       
	    \multicolumn{1}{l|}{Core 2 [\%]}&96.0  &93.8   &92.7   &99.5   &83.1   &94.1   &74.9   &70.4   &&&&                              &67.6     &49,9      &58.7      &66.6    \\
	    CL update  &4.17  &4.26   &4.31   &8.04   &7.21   &8.49   &13.34  &22.72  &&&&                              &29.60    &72.04     &88.61     &108.15  \\
	    GB/s       &43.5  &42.5   &84.1   &67.7   &25.1   &42.7   &40.7   &31.9   &&&&                              &6.1      &5.0       &6.1       &6.7     \\
	    eff. GB/s  &-     &-      &-      &-      &-      &21.3   &27.2   &23.9   &&&&                              &-        &2.5       &4.1       &5.0     \\
	    \hline                                                                                                                                                      
	    \multicolumn{1}{l|}{Nehalem [\%]}&97.1  &95.3    &94.1   &96.0   &83.5   &120.9  &91.4   &91.7   &95.3   &121.4    &103.9   &96.3    &106.8    &142.2     &123      &119.4  \\
	    CL update   &4.12  &4.20    &4.26   &8.34   &7.18   &6.61   &10.94  &17.45  &8.39   &9.88    &15.4   &24.91   &14.02    &18.27     &29.25    &42.72  \\
	    GB/s        &41.3  &40.5    &79.8   &61.2   &23.7   &51.5   &46.7   &39.0   &20.3   &34.4    &33.2   &27.3    &12.1     &18.6      &17.4     &15.9   \\
	    eff. GB/s   &-     &-       &-      &-      &-      &25.7   &31.1   &29.3   &-      &17.2    &22.1   &20.5    &-        &9.3       &11.6     &11.9   \\
	    \hline                                                                                                                                                    
	    \multicolumn{1}{l|}{Shanghai [\%]}&88.3  &95.3    &97.1   &85.0   &74.5   &55.1   &80.6   &78.5   &48.9   &54.9    &50.6   &49.5    &75.4    &75.6     &80.8    &80.6   \\
	    CL update    &2.27  &4.20    &6.18   &9.41   &8.05   &13.58  &17.36  &25.47  &16.36  &18.20   &35.53  &50.7    &23.86   &42.32    &61.89    &84.32   \\
	    GB/s         &67.9  &36.7    &49.9   &49.2   &19.2   &22.7   &35.6   &36.4   &9.4    &16.9    &17.4   &18.1    &6.5     &7.3      &7.4      &6.9    \\
	    eff. GB/s    &-     &-       &-      &-      &-      &11.4   &17.8   &18.2   &-      &8.5     &8.7    &9.0     &-       &3.6      &4.9      &5.5     \\
	\end{tabular}
\end{table}

As for main memory access, one must distinguish between the classic frontside
bus concept as used with all Core 2 designs, and the newer architectures with
on-chip memory controller. The former has much larger overhead, which is why
Core 2 shows mediocre efficiencies of around 60\,\%\@. The AMD Shanghai, on the
other hand, reaches around 80\,\% on all benchmarks. The Core i7 shows results
better than the theoretical prediction.  This can be caused either by a potential
overlap between contributions or by the inaccurate store model.
%
\subsection{Multi-Threaded Stream Triad Performance}
%
An analytical extension of the model to multi-threading is beyond the scope of
this work and would involve additional analysis of the cache subsystems and
threaded execution. However, as bandwidth scalability of shared caches on
multi-core processors is extremely important for parallel code optimization, we
determine the basic scaling behavior of the cache subsystem using
multi-threaded stream triad bandwidth tests. The Core 2 Quad processor used
here comprises two dual-core chips in a common package (socket), each with a
shared L2 cache. On the other two architectures each core has a private L2
cache and all cores share an L3 cache. 
For our tests, threading was implemented based on the POSIX threading
library, and threads were explicitly pinned to
exclusive cores. Pinning was done with a wrapper library overloading the
\verb+pthread_create()+ function.

The measurements for four threads on the Core 2 architecture did not produce
reasonable results due to the large and varying barrier overhead. The shared
L2 cache for the Core 2 scales to two cores (Table \ref{tab:multi_stream}).
This is possible by interleaving the reload and the execution of the
instructions in L1 cache between the two cores, as described earlier. The same
is valid for the shared L3 caches on the other two architectures. The L3 cache
on the Intel Core i7 scales up to two threads but shows strong saturation with
four threads. 

On all architectures, a single thread is not able to saturate the memory bus
completely. This can be explained by the assumption that also for main memory
access only a part of the runtime is usable for data transfer. Of course, this
effect becomes more important if data transfer time from main memory is short
compared to the time it takes to update and transfer the data on the processor
chip. Another reason may be an insufficient number of outstanding prefetch
operations per core, so that multiple cores are needed to hide main memory
latency completely.
\begin{table}[tb]
    \caption{Threaded stream triad performance}
    \label{tab:multi_stream}\addtolength{\tabcolsep}{1mm}
    \centering
	\begin{tabular}{ll|ccccc}
	                   &Threads   &L1      &L2       &L3     &Memory \\
	    \hline
	    Core 2 [GB/s]  &1         &66.1    &23.7    &-       &4.9   \\
	                   &2 (shared)&134.1   &46.9    &-       &5.0   \\
	                   &4         &-       &-       &-       &5.3   \\
	    \hline                                                              
	    Nehalem [GB/s]&1          &61.1    &29.0    &20.5    &11.9   \\
	                  &2          &122.1   &55.9    &39.8    &14.8  \\
	                  &4          &247.7   &113.3   &51.3    &16.1  \\
	    \hline                                                                            
	    Shanghai [GB/s]&1         &49.2    &17.7    &9.1     &5.5   \\
	                  &2          &49.1    &35.3    &19.5    &7.1   \\
	                  &4          &187.0   &70.7    &36.9    &7.9   \\
	\end{tabular}
\end{table}
%
\section{Conclusion}
\label{sec:conclusion}
%
The proposed model introduces a systematic approach to understand the
performance of bandwidth-limited loop kernels, especially for in-cache
situations. Using elementary data transfer operations we have demonstrated the
basic application of the model on three modern quad-core architectures. The
model explains the bandwidth results for different cache levels and shows that
performance for bandwidth-limited kernels depends crucially on the runtime
behavior of the instruction code in L1 cache. This work proposes a systematic
approach to understand the performance of bandwidth-limited algorithms. It does
not claim to give a comprehensive explanation for every aspect in the behavior
of the three covered architectures. 

Work in progress involves application of the model to more
relevant algorithms like, e.g., the Jacobi or Gauss-Seidel smoothers.
Future work will include verification and refinements for the
architectures under consideration. An important component is to fully
extend it to multi-threaded applications. Another possible application
of the model is to quantitatively measure the influence of hardware
prefetchers by selectively disabling them.
%
%

\bibliographystyle{splncs}
\end{document}